\documentclass[10pt]{article}
\usepackage{epsf}

\usepackage{epsfig,graphics}
\usepackage {graphicx}
\usepackage {epsfig}
\usepackage {subfigure}
\usepackage {tabularx} 
\usepackage{rotate}	
\usepackage{slashed}

\usepackage{amsmath}
\usepackage{amsfonts}
\usepackage{amssymb}
\usepackage{graphicx}
\usepackage{cite}

\usepackage{fancyhdr}
\usepackage{hyperref}

\newcommand{\bmat}{\left(\begin{array}}
	\newcommand{\emat}{\end{array}\right)}

\def\yzero{\smash{\hbox{$y\kern-4pt\raise1pt\hbox{${}^\circ$}$}}}

\def\beq{\begin{equation}}
\def\eeq{\end{equation}}
\def\beqa{\begin{eqnarray}}
\def\eeqa{\end{eqnarray}}

\def\-{\hphantom{-}}

\def\s2{\frac{1}{\sqrt2}}

\def\beq{\begin{equation}}
\def\eeq{\end{equation}}
\def\beqa{\begin{eqnarray}}
\def\eeqa{\end{eqnarray}}

\def\IF{\relax{\rm I\kern-.18em F}}
\def\II{\relax{\rm I\kern-.18em I}}
\def\IP{\relax{\rm I\kern-.18em P}}
\def\IC{\relax\hbox{\kern.25em$\inbar\kern-.3em{\rm C}$}}
\def\IR{\relax{\rm I\kern-.18em R}}

\def\Dsl{\,\raise.15ex\hbox{/}\mkern-13.5mu D} 
\def\IZ{Z\kern-.4em  Z}

\catcode`\@=11   
\newdimen\@rotdimen
\newbox\@rotbox  

\def\@vspec#1{\special{ps:#1}}
\def\@rotstart#1{\@vspec{gsave currentpoint currentpoint translate
		#1 neg exch neg exch translate}}
\def\@rotfinish{\@vspec{currentpoint grestore moveto}}
\def\@rotr#1{\@rotdimen=\ht#1\advance\@rotdimen by\dp#1%
	\hbox to\@rotdimen{\hskip\ht#1\vbox to\wd#1{\@rotstart{90 rotate}%
			\box#1\vss}\hss}\@rotfinish}
\def\@rotl#1{\@rotdimen=\ht#1\advance\@rotdimen by\dp#1%
	\hbox to\@rotdimen{\vbox to\wd#1{\vskip\wd#1\@rotstart{270 rotate}%
			\box#1\vss}\hss}\@rotfinish}%
\def\@rotu#1{\@rotdimen=\ht#1\advance\@rotdimen by\dp#1%
	\hbox to\wd#1{\hskip\wd#1\vbox to\@rotdimen{\vskip\@rotdimen
			\@rotstart{-1 dup scale}\box#1\vss}\hss}\@rotfinish}%
\def\@rotf#1{\hbox to\wd#1{\hskip\wd#1\@rotstart{-1 1 scale}%
		\box#1\hss}\@rotfinish}%
\def\rotate{\@ifnextchar[{\@rotate}{\@rotate[l]}}
\def\@rotate[#1]#2{\setbox\@rotbox=\hbox{#2}\@nameuse{@rot#1}\@rotbox}

\catcode`\@=12

\topmargin
-1.5cm
\textwidth
15.5cm
\textheight
23.5cm
\oddsidemargin
0.7cm
\evensidemargin
0.7cm

\begin{document}

	\makeatletter
	\@addtoreset{equation}{section}
	\makeatother
	\renewcommand{\theequation}{\thesection.\arabic{equation}}
	\pagestyle{empty}
	\vspace{-0.2cm}
	\rightline{ IFT-UAM/CSIC-19-25}
	\vspace{1.2cm}
	\begin{center}
		
		
		\LARGE{ A Strong Scalar Weak Gravity Conjecture \\
			and Some Implications\\ [13mm]}
		
		\large{Eduardo Gonzalo  and Luis E. Ib\'a\~nez \\[6mm]}
		\small{
			Departamento de F\'{\i}sica Te\'orica
			and Instituto de F\'{\i}sica Te\'orica UAM/CSIC,\\[-0.3em]
			Universidad Aut\'onoma de Madrid,
			Cantoblanco, 28049 Madrid, Spain 
			\\[8mm]}
		\small{\bf Abstract} \\[6mm]
	\end{center}
	\begin{center}
		\begin{minipage}[h]{15.22cm}
				We propose a new version of the scalar Weak Gravity Conjecture (WGC) which would apply to 
				any scalar field  coupled to quantum gravity. For a single scalar it is given by the differential constraint
				$(V'')^2\leq  (2V'''^2- V''V'''')M_{\text{p}}^2$,
				 where $V$ is the scalar potential. 
				 It corresponds to the statement 
				that self-interactions of a scalar  must be stronger than gravity for any value of the scalar field. 
				We find that the solutions which saturate the bound correspond to towers of extremal states
				with mass $m^2(\phi)=m_0^2/((R/m)^2+1/(nR)^2)$, with $R^2=e^\phi$, consistent  with the emergence of an extra dimension
				at large or small $R$ and the existence of extended objects (strings). These states act as WGC states for the scalar $\phi$.
				It  is also consistent with the distance swampland conjecture with a built-in  duality symmetry.
				All of this is remarkable since neither extra dimensions nor string theory are put in the theory from the beginning, but they emerge.
				This is quite analogous  to how the 11-th dimension appears in M-theory from towers of Type IIA solitonic $D0$-branes.
				From this constraint one can derive several swampland conjectures from a single principle. In particular 
				one finds that an axion potential is only consistent  if  $f\leq M_{\text{p}}$, recovering a result
				already conjectured from other arguments.
								 The conjecture has far reaching consequences and applies to several
				interesting physical systems: i) Among chaotic inflation potentials only those asymptotically linear may survive. ii) 
				If applied to the radion of the circle compactification of the Standard Model to 3D with Dirac neutrinos, the
				constraint implies that the 4D cosmological constant scale must be larger than the mass of the lightest neutrino,
				which must be in normal hierarchy. It also puts a constraint on the EW scale, potentially explaining the hierarchy problem.
				This recovers and improves results already obtained by applying the AdS swampland conjecture, but
				in a way which is independent from UV physics.  iii)
				It also constraints simplest moduli fixing string models. The simplest KKLT model is compatible with
				the constraints but the latter may be relevant for some choices of parameters.

		\end{minipage}
	\end{center}
	\newpage
	\setcounter{page}{1}
	\pagestyle{plain}
	\renewcommand{\thefootnote}{\arabic{footnote}}
	\setcounter{footnote}{0}
	

	
	\tableofcontents
	
	\section{A Scalar Weak Gravity Conjecture}

The Weak Gravity Conjecture (WGC) was first formulated in \cite{swampland,WGC,distance}, see \cite{review} for a recent review and
more references. 
The most widely studied WGC example is the case of a 
$U(1)$ gauge boson coupled to gravity. It states that there must always exist a charged particle with mass m and charge q such that 
$m\leq gqM_{\text{p}}$ in the theory. Arguments based on extremal charged back-holes and string theory examples give solid
support to this conjecture, which has been generalized to multiple $U(1)$'s as well as antisymmetric tensor couplings in supergravity and string theory,
see ref.\cite{WGC1,WGC2,WGC3,timo} for some recent reverences and \cite{review,vafafederico} for an introduction.

There are however two options concerning what  is the most important physical principle underlying  the WGC, either 1) It is something primarily related 
 to black-holes and their stability or rather 2) It is the general principle of gravity being the weakest force which is the crucial point.
If the second is true,  the consequences would be paramount,  since there are many physical instances in which interactions may potentially be
weaker than gravity without black-holes playing (at least apparently)  any role. In the present paper we want to argue that insisting in gravity
being always and in any circumstance the weakest force, may have very important implications if applied to scalar particles. 

In this paper we put forward the proposal of a {\it Strong Scalar WGC} which is defined by eq.(\ref{ours}), corresponding to
the statement that the self-interactions of a scalar must be stronger than the gravitational force for all values of the
scalar field. This must be true for any scalar in the theory, and not only for a particular set of WGC states. The 
extremal version of the equation yields a surprise: the solutions are compatible with towers of momenta and winding of
an emerging dimension. Those towers become massless for $|\phi|\rightarrow \infty$, in agreement, with expectations from the 
swampland distance conjecture \cite{distance,review,vafafederico}. We interpret these towers as massive solitonic states which appear  playing the role of WGC states.
This structure is analogous to the behaviour in string theory in which towers of solitonic states (D-branes)  become massless or tensionless
for large fields.
Thus the simplest theory you can think of, a scalar coupled to quantum gravity,  secretly contains several features of string theory: emerging extra dimensions,
winding strings and duality.

\subsection{The scalar weak gravity conjecture}

The WGC case for purely scalar interactions is not obvious, since clear arguments based on blackhole physics  are lacking. 
Still it has been argued that a variant of the WGC applies to axions with masses replaced by instanton actions and gauge couplings 
replaced by $1/f$, with $f$ the axion decay constant.
 For axions the  corresponding bound is \cite{swampland,WGC,WGC1,review}
 \beq 
 S_{inst}\ \leq \frac {1}{f}M_{\text{p}} \ ,
 \eeq
 where $S_{inst}$ is the instanton action.   For the theory to be within control one asks for $S_{inst}\geq 1$, leading to the 
 constraint $f\leq M_{\text{p}}$. This is relevant for models of natural inflation in which values for $f$ larger than $M_{\text{p}}$ are in general required
 in order to get appropriate inflation.

Palti formulated a   first version of a Scalar Weak Gravity Conjecture (SWGC) in the following terms \cite{Palti} (see also \cite{timoscalar} and \cite{Lust,review}). We consider a particle  H with mass $m$ which is coupled to 
a light scalar $\phi$ with a trilinear coupling proportional to  $\mu=\partial_\phi m$.  Then the conjecture is
that,  as $m_\phi \rightarrow 0$, 
\beq
 (\partial_\phi m)^2\geq \frac {m^2}{M_{\text{p}}^2} \ .
 \eeq
The statement is that the force mediated by $\phi$ must be stronger than the gravitational force and  $m^2(\phi)$ is considered as a function of $\phi$ so that
$m^2=V''$.  So the above expression may be
written as 
\beq
(V''')^2\geq \frac {(V'')^2}{M_{\text{p}}^2} \ .
\label{palti}
\eeq
Here the particle $H$ acts as a WGC particle in the sense that it is there to guarantee that
there  is at least one  particle with interactions stronger than gravity.  The philosophy sounds similar to the WGC for gauge $U(1)$ interactions. 
However, both situations are apparently very different. In particular the scalar has no charge which could create  a blackhole stability problem as with
charged fields under a $U(1)$,  and the generalization is not obvious.

 \begin{figure}
	\centering{}\includegraphics[scale=0.3]{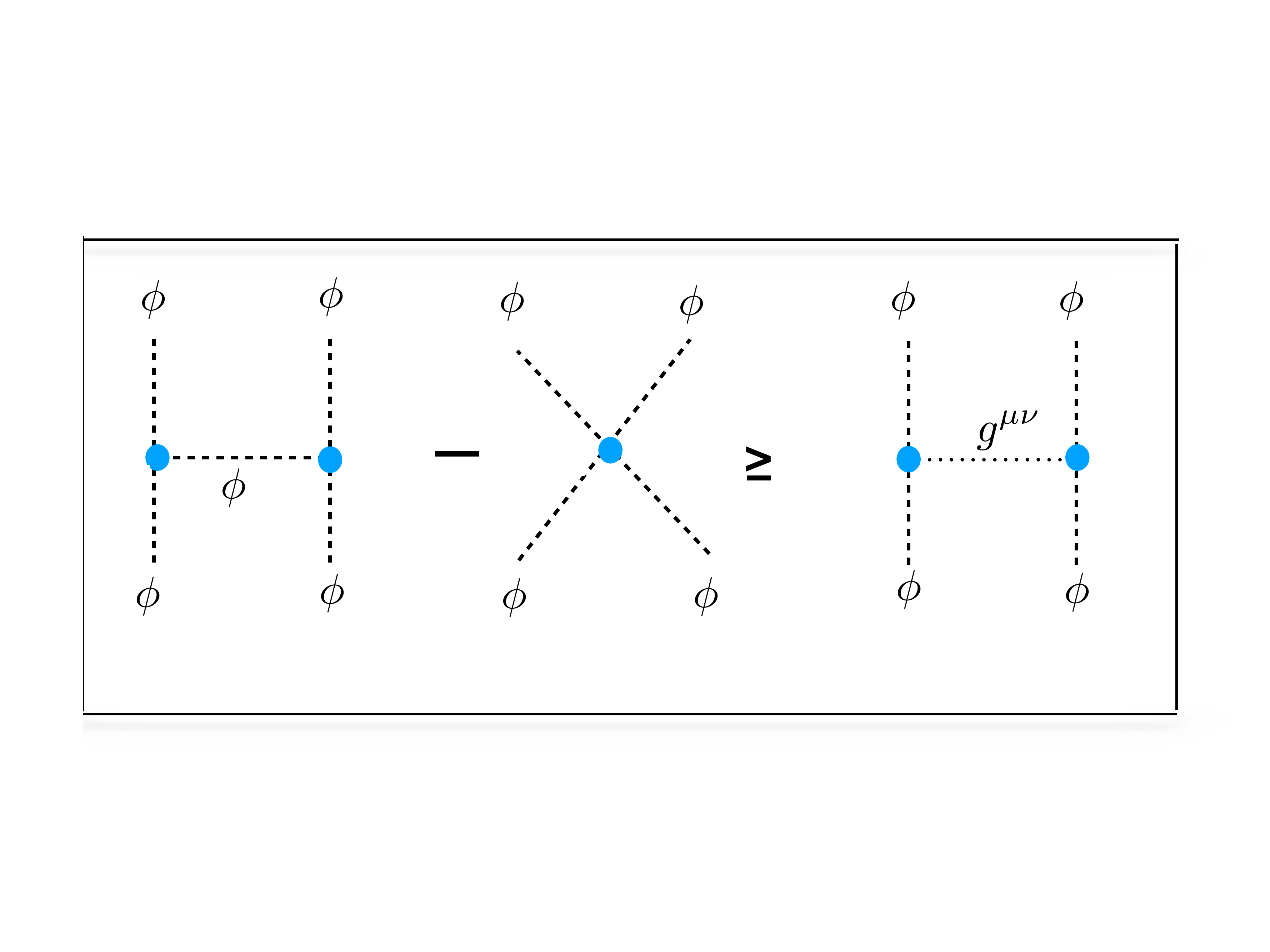}
	\caption{ Pictorial representation of the the Strong Scalar Weak Gravity Conjecture}
 \label{higgsfinal}
\end{figure}

\subsection{A Strong scalar WGC}

As formulated by Palti, the above bound does not apply to any scalar, but only to WGC scalars which interact with a scalar $\phi$ and whose mass
is a function  of $\phi$. That is for example the case of some string particle (like a lightest KK mode) which depends on the moduli of a compactification.
The conjecture does not apply as it stands to the fields $\phi$ themselves.
With the above expression one finds that the marginal situation for the mass of the WGC scalar $H$ occurs for
\beq
m^2\  = \ m_0^2e^{\pm \phi/M_{\text{p}}} \ .
\eeq
This gives the expected behaviour which appears in the distance conjecture at large $\phi$ \cite{distance} (see also \cite{paltidistance}).  So this scalar $H$ could be like the
first member of a tower. 
However this exponential behaviour is at odds with the properties of axions, whose potential 
is periodic and hence inconsistent with eq.(\ref{palti}). Also, the exponential must be $e^{-\phi/M_{\text{p}}}$ for large $\phi$ but
one must change to $e^{\phi/M_{\text{p}}}$ for $\phi \rightarrow - \infty$, and there is no single function which includes both behaviours simultaneously.
We  propose that the above formulated SWGC needs to be modified in such a way that both issues may be circumvented.  Furthermore we will 
insist that our new SWGC applies to any scalar in the theory. The latter possibility was termed {\it Super SWGC} in \cite{Palti}.

We propose the formulation of a  Strong version of a Scalar Weak Gravity Conjecture (SSWGC) 
for the case of a single scalar as follows:

\vspace{0.3cm}

{\it i) The potential of any canonically normalized real scalar  $V(\phi)$ in the theory must verify for any value of the field the constraint:}

\beq
 \boxed{
  2(V''')^2  -\ V''V''''\ \geq \ \frac {(V'')^2}{M_{\text{p}}^2}, \ 
  } 
  \label{ours}
\eeq
\vspace{0.3cm}

with primes denoting derivation with respect to $\phi$. Compared to eq.(\ref{palti}) here there is a new term
 $V^{\prime \prime \prime \prime}$ associated to the quartic interaction of the scalars. Such a term
was not present  in the SWGC bound in the previous paragraph because such an interaction among the $H$ fields is 
not mediated by $\phi$ and hence it should not be included.  In our case it  is different because our condition applies to any scalar,
including massive mediators. In our conjecture there are no additional WGC H scalars present in the spectrum to verify a
WGC. Rather the states playing that role will be towers of extremal  collective objects, as described in section (\ref{clave}).

Eq.(\ref{ours}) looks like a condition which imposes that the strength of a scalar interaction must always 
be stronger than gravity.
The presence of the quartic term is crucial to
obtain the required consistency  for the axion potential and is also justified a posteriori by the results in section  (\ref{clave}).
In fact the factors and signs of the terms in the left are crucial in order to obtain the nice results in that section.
We come short of having a Feynman graph explanation for the above differential constraint.
One can motivate this expression by considering the short distance behaviour of the potential between two scalars,
see fig.(\ref{higgsfinal}).
At short distances the
first term comes from the exchange of the scalar $\phi$, which has the same attractive behaviour than the Newtonian term, $V\simeq -1/r$.
The  second term includes a direct quartic piece, which is repulsive and proportional to a Dirac delta, hence an UV contribution.
On the other hand, in the IR regime, due to the trilinear coupling being super-renormalizable,  the first term gives rise to an effective contact
 term which is attractive.
Thus one cannot factor out a universal
distance dependence. In fact  eq.(\ref{ours}) seems to encapsulate mixed UV and IR effects. This is perhaps not surprising considering the results in section
(\ref{clave}). The presence of the quartic terms is crucial for the presence  of winding states and duality in the emergent dimension.

Before proceeding, some comments about simple potentials are in order:
\begin{itemize}
\item A linear potential $V=a\phi+c$ always verify our  SSWGC. This means that the value of $|\phi|$ is unconstrained and
may be trans-Planckian with no inconsistency.
\item  A pure quadratic potential $V=m^2\phi^2$ is special. In this case the condition is violated  for any value of $\phi$ with
$m^2>0$. This may be interpreted as a condition that {\it  forbids the existence of massive scalars with no interaction other than
gravity}. It reminds the $U(1)$ WGC which also states that gauge bosons must have at least one charged particle to interact with.

 \item For a purely cubic(quartic)  potential $V=\mu\phi^3(V=\lambda\phi^4)$ the conditions are fulfilled only if $|\phi|\leq \sqrt{2}M_{\text{p}}(|\phi|\leq \sqrt{6}M_{\text{p}})$. 
\end{itemize}

An exponential potential of the simple runaway  form $V=\exp(\pm \alpha\phi)$  passes the test as long as $|\alpha|\geq 1/M_{\text{p}}^2$.
Note also that the constraint is insensitive to $V'$ and $V$ themselves so insensitive to whether the theory is
in dS or AdS and the conditions for minima. So the constraints here discussed seem unrelated to the dS conditions of ref.\cite{dS1,dS2,krishnan,dS3,masdS}.
In particular our condition is compatible but independent from the dS conjecture.

The above constraint may be easily generalized to the case of several scalars fields.

\subsection{A first test: the axion potential}

If the SSWGC applies to any scalars, it should apply also to axions and their periodic potentials which we know appear in string theory
whenever an axion-like scalar couples to a non-Abelian gauge group. So one may consider the axion example as a test for the conjecture.
The leading instanton contribution to the axion potential has the form
\beq
V\ =\   -\ \cos (\phi/f)) \ .
\eeq
In this case the SSWGC gives
\beq
\frac {1}{f^6}\left(2 \sin^2(\phi/M_{\text{p}})\ +\cos^2(\phi/M_{\text{p}})\right) \ \geq \ \frac{ \cos^2(\phi/M_{\text{p}})}{f^4M_{\text{p}}^2}\ .
\eeq
Here we have constrained ourselves to the region in which $V''\geq 0$ in which the leading cosine instanton term is expected to dominate.
The above expression yields 
\beq
f^2\ \leq \  M_{\text{p}}^2(1+2 \tan^2(\phi/M_{\text{p}}))
\eeq
and, since the bound must be true for all $\phi$,  one obtains $f\leq M_{\text{p}}$. So we see we can derive from the SSWGC the condition 
that the
decay constant  f of an axion  cannot exceed $M_{\text{p}}$
\cite{swampland,WGC,WGC1}. In the present case this comes about because otherwise the 
scalar interactions would be weaker than gravitation.

\subsection{ The extremal equation: towers of states and an emerging dimension}
\label{clave}

We can consider the {\it extremal} case for a single scalar in which the scalar interactions equal the gravitational one.
Then the constraint may be written as a differential equation on the field dependent mass $m^2(\phi)$:
\beq
2((m^{2})^{ \prime})^2 \ -\  m^2( (m^{2})^{ \prime \prime})-  \frac {m^4}{M_{p}^2}\ =\ 0 \ .
\label{guay}
\eeq
One obtains the extremal solutions for $m^2$ (with $\phi$ in $M_{\text{p}}$ units):
\beq
 m^2(\phi)\ = \  \frac {Ae^\phi}{Be^{2\phi}+1} \ .
 \eeq 
 For this to be a solution one must have $B\geq0$. Concerning $A$, we chose it positive (otherwise $m^2$ would always be negative for all $\phi$).
 Defining a field $R=e^{\phi/2}$, with kinetic metric $2(dR/R)^2$
 one can rewrite the above expression in the more suggestive way
 \beq
 m^2\ =\ \frac {m_0^2}{1/(NR)^2\ +\ (R/M)^2} \ .
 \eeq
 For $N,M\not=0$ one can also  write 
 \beq
 m^2\ =\  m_0^2\frac { (NM)^2}{{\cal M}_{N,M}^2}\ ,\  {\cal M}_{N,M}^2=N^2R^2 +\frac {M^2}{R^2} \ .
 \label{masilla}
 \eeq
 Here ${\cal M}_{N,M}$ looks like the spectrum of a string compactification in a circle,
 with the duality invariance
 \beq
 R\ \longleftrightarrow 1/R \ ;\  M\ \longleftrightarrow \ N \ .
 \eeq
  Note also that for large(small)  $R$ one gets the limits:
 \beq 
 m_{\phi \rightarrow \infty}^2 \longrightarrow\  {m_0^2}\ M^2e^{-\phi}  \ ;\ m_{\phi \rightarrow -\infty}^2 \longrightarrow\  {m_0^2}\ N^2e^{\phi}
 \eeq
 For integer $N,M$ this has the structure of towers of winding and momenta becoming light as the scalar $\phi$ goes to infinity. 
 Our interpretation is that these towers are the WGC scalars which are required so that  gravity keeps on being the weakest force when
 $|\phi|$ goes to infinity. If $\phi$ is identified with a modulus,
 this is precisely the statement   in the swampland distance conjecture
 \cite{distance,paltidistance,irene,thomas}.  Notice also that the extremal solutions know about winding and hence about 
 string theory. This is in agreement with the argument in \cite{vafafederico} that the distance conjecture requires the existence of extended objects.
 Thus    towers of quantized 
 momenta and winding  from a  5D string compactified on a circle 
 saturate the 4D Strong scalar WGC.
 This is remarkable, since there is no explicit circle compactification nor 
 winding strings in the original differential equation. 
 A dimension of radius $R$ emerges from the condition of the Strong SWGC conjecture. If this is the case, there should also be an
 emerging graviphoton under which the momenta are charged, justifying a posteriori choosing $N,M$ integers. 
Note finally that obviously one could rather identify    $e^\phi$ with a {\it gauge} coupling  $g$ of the complete theory, in which case  as
 $g^2\rightarrow 0$  a tower of states become massless to preclude the presence of global symmetries in the  effective theory.

The argument goes also in the opposite direction. Consider a 4D theory obtained upon compactification of a 5D string theory on a circle.
Then the masses of the particles in the KK and winding towers depend on the radion in such a way that their potential verifies the
differential equation (\ref{guay}). This gives support to the proposed conjecture and the equation.

Given the above discussion, we propose a second conjecture:

\vspace{0.3cm}
{\it ii) In the system of a single canonically normalized scalar field $\phi$ coupled  to quantum gravity,  there  are extremal massive states which have a structure corresponding to 
momenta and winding states of a string compactified in a circle of radius $R^2=e^\phi$, corresponding  to an emergent dimension. Those states come in WGC towers which become 
massless as $|\phi|\rightarrow \infty$.}
\vspace{0.3cm}

Note that the structure is analogous to how the 11-th dimension appears from Type IIA string theory at strong coupling. The analogue of the above 
extremal states are the towers of $D0$-branes of string theory building up the KK modes of the 11-th dimension. Thus the simplest system one can think of 
with a single scalar coupled to quantum gravity secretly has several features characteristic of string theory.

\subsection{Extremal potentials}

 Independently of the existence of these towers of states, it is interesting the question of whether one can write down potentials 
 saturating the bound. 
By integrating  $V''$ one can obtain general forms of potentials verifying  the extremal case in which the 
inequality is saturated. One finds solutions of the general form
\beq
V'\ =\ \frac { A}{{\sqrt{B}}}\  \tan^{-1}(\sqrt{B}e^\phi)\ +\ C \ .
\eeq
Further integration yields for the extremal potential 
\beq
V(\phi)_{extr}\ =\ 
\frac {iA}{2\sqrt{B}}\left( \text{Li}_2(-i\sqrt{B}e^\phi)\ -\ \text{Li}_2(i\sqrt{B}e^\phi)\right)\ + \ C\phi \ +\ D \ .
\eeq
where $\text{Li}_2$ is the dilogarithm function and $i=\sqrt{-1}$. In spite of its complex appearance the potential is real (for real constants). One can see that the first term
grows linear with $\phi$ as  $\phi\rightarrow +\infty$ and is damped exponentially as $\phi \rightarrow - \infty$. This asymptotic linear behavior is in agreement with our
comment above that linear potentials always satisfy the differential constraint.

\begin{figure}
	\centering{}\includegraphics[scale=0.5]{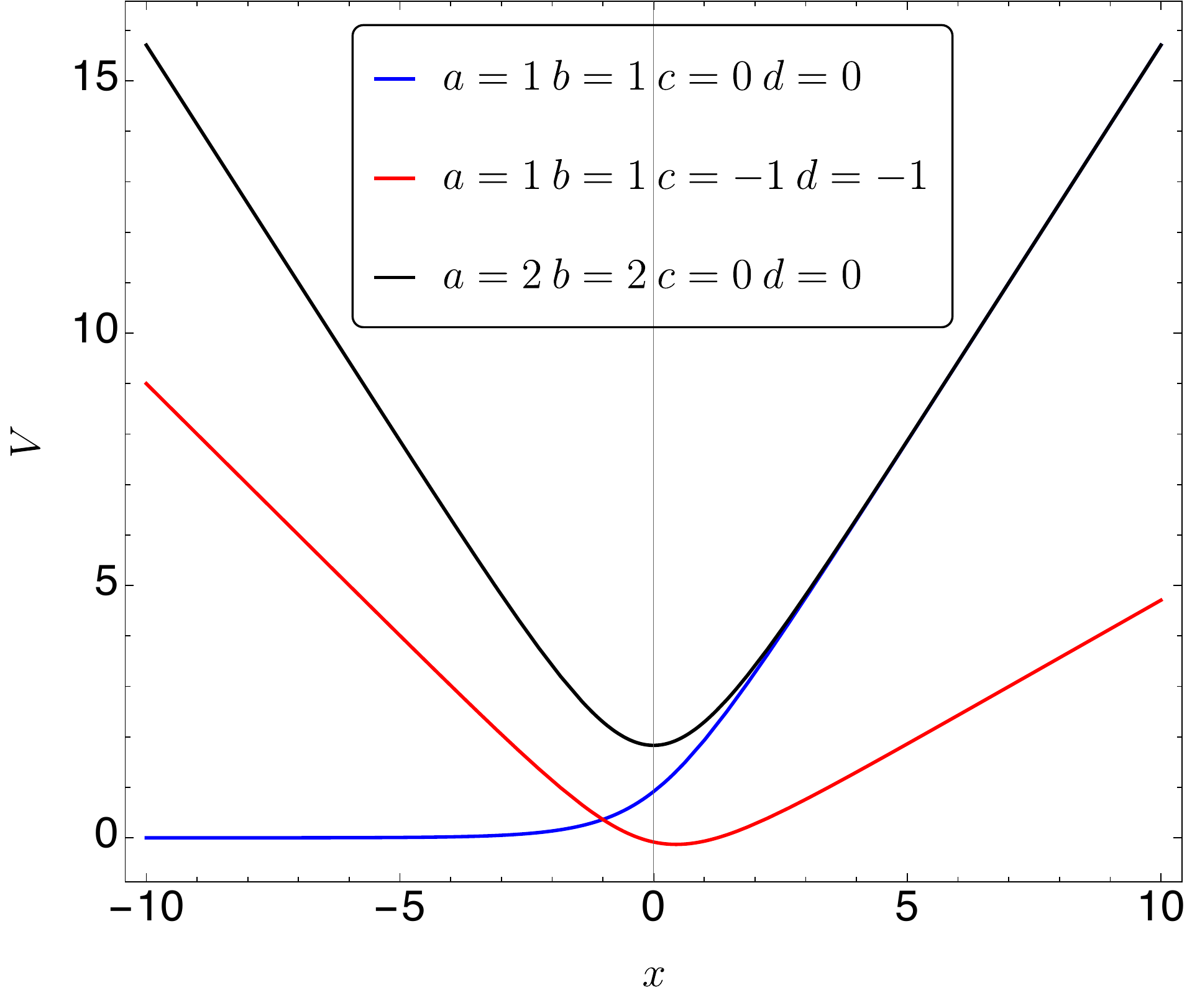}
	\caption{Examples of extremal potentials.}
 \label{extremal}
\end{figure}

This class of potentials depend on 4 real constants $A,B,C,D$.  For $A=0$  one just gets straight lines. For other choices one may get also runaway 
potentials as well as minima which may be dS, AdS or Minkowski depending on the choice for D. Some simple interesting cases are as follows:

\begin{itemize} 

\item  i) $A=B=1;  C=D=0$, fig.\ref{extremal} in blue.
The potential is linear at large $\phi$ and exponentially decreasing for large negative $\phi$. So this is an example of a runaway potential.

\item  ii) $A=B=1; C=-1$,$D=0$, fig.\ref{extremal} in red.
The potential shows a minimum and behaves linearly for $|\phi|>M_{\text{p}}$. This minimum may be in dS or AdS depending on the choice for $D$.

\item

iii) Those two cases saturate the bound but are not duality invariant.
 If  however one insists in a duality invariance $\phi \leftrightarrow -\phi$ one has 
\beq 
V(\phi)_{extr}\ =\ 
\frac {iA}{2\sqrt{B}}\left(\text{Li}_2(-i\sqrt{B}e^\phi)\ -\ \text{Li}_2(i\sqrt{B}e^\phi)\right)\ + \  (\phi \leftrightarrow -\phi) \ 
\eeq
The potential is then symmetric with a minimum at $\phi=0$. For $A=2, B=1$, $C=D=0$ this is depicted in fig.\ref{extremal} in black.
This class of potentials is interesting in its own right and may have interesting physical applications e.g. in inflation. In particular,
given its asymptotic linear behaviour, it should give rise to a variation of linear inflation.

\end{itemize}

\subsection{Constraints on some simple potentials}

It is interesting to see what are the constraints for a scalar potential of the form
\beq
V(\phi)\ =\ \frac {1}{2}m^2\phi^2\ +\ \frac {1}{4\!}\lambda \phi^4 \ .
\eeq
It is easy to see one finds the constraint
\beq
\lambda(\frac {3}{2}\phi^2\ -\ m^2)\ \geq \ \frac {1}{M_p^2}(m^2+\frac {\lambda}{2} \phi^2)^2 \ .
\eeq
Note that the term in the right is strongly suppressed by the $M_p^{-2}$ factor, so that in practice (for $\phi^2\ll M_p^2$) the constraint 
amounts to the left hand side being positive. This is automatic for $m^2<0$ and $\lambda$ positive. This is similar to the situation in the SM.
On the other hand for $m^2>0$ the constraint is only obeyed for $\phi^2>(2/3)m^2$. 

A simple class of SUSY superpotentials is the exponential one, 
$W=e^{-\alpha M}$, with $M$ a canonically normalized complex scalar.
The condition may be written as
\beq
2(V_{MM^*M^*})(V_{M^*MM})\ -\ (V_{MM^*})(V_{MMM^*M^*})\ \geq \ \frac {(V_{MM^*})^2}{M_{\text{p}}^2} \ .
\eeq
It is easy to check that this exponential superpotential leads to a potential passing the  SSWGC constraint as long as 
$|\alpha |\geq 1/M_{\text{p}}^2$, for any value of $M$. One can also test a cubic superpotential which may appear in e.g. in string flux compactifications, i.e.
\beq
W(T)\ =\ \frac {m}{2}T^2\ +\ \frac {\lambda}{6}T^3 \ .
\eeq
The differential inequation yields
\beq
 \lambda^2 \ \geq \ \frac {|m+\lambda T|^2}{M_{\text{p}}^2} \ .
 \eeq
One sees that trans-Planckian trips of $T$ would in this case violate the bound. And also the scalar interaction  coupling is bounded
below by $\lambda^2 \geq \frac {m^2}{M_{\text{p}}^2}$. This is consistent with the idea that gravity must be the weakest force. 
Let us comment that in fact instead of the global SUSY potential one should have used the $N=1$ supergravity potential. 
However this does not modify the result because the additional terms in the potential have an extra Planck mass supression.

\section{ Applications}

The above introduced Strong SWGC may have an important impact whenever there is some Planck suppressed  scalar interaction with the risk of
becoming weaker than gravity. Here we list four  important applications leaving a more detailed account for a future publication.

\subsection{Inflation}

\begin{figure}
	\centering{}
	\includegraphics[scale=0.38]{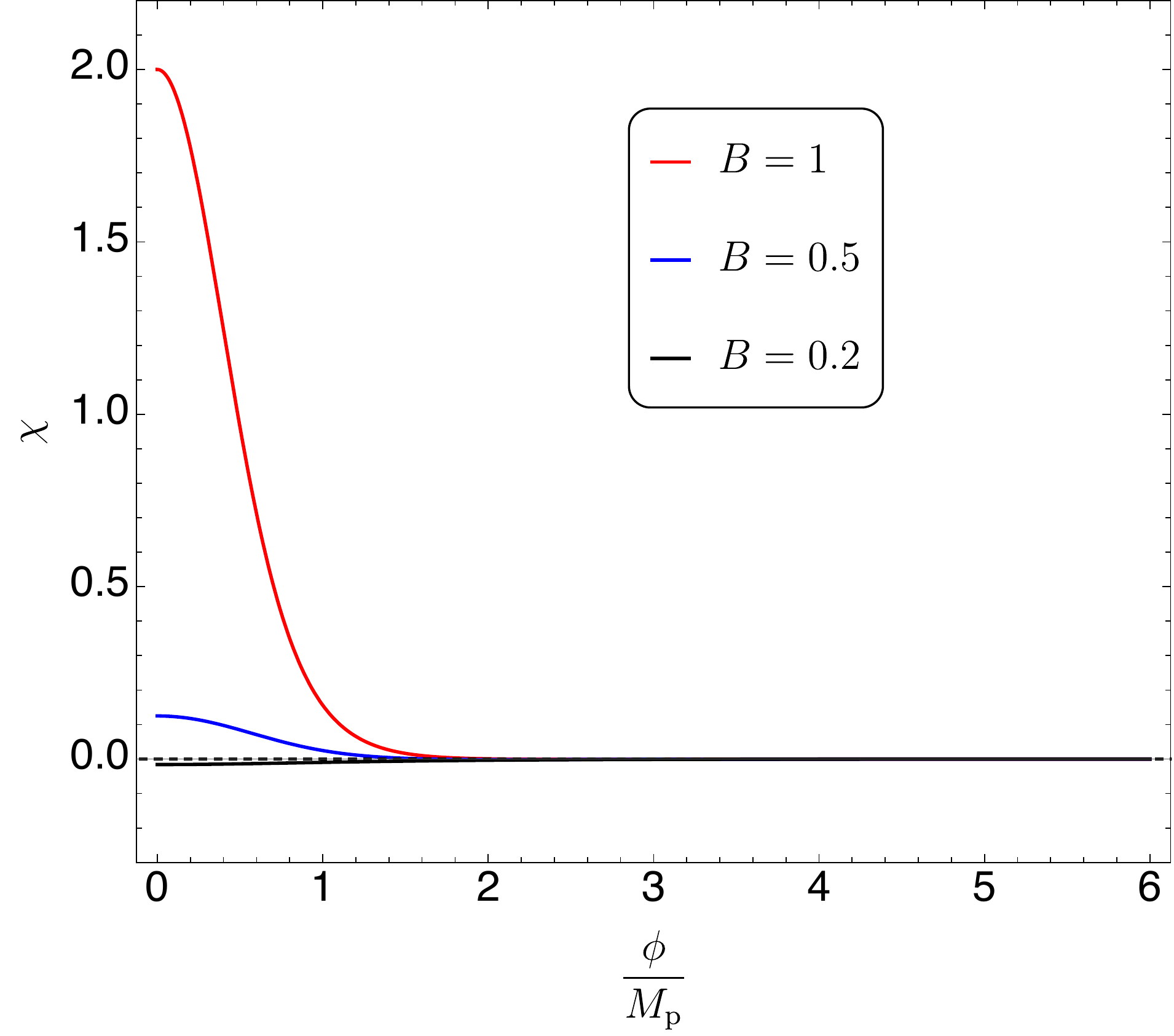} 	\includegraphics[scale=0.38]{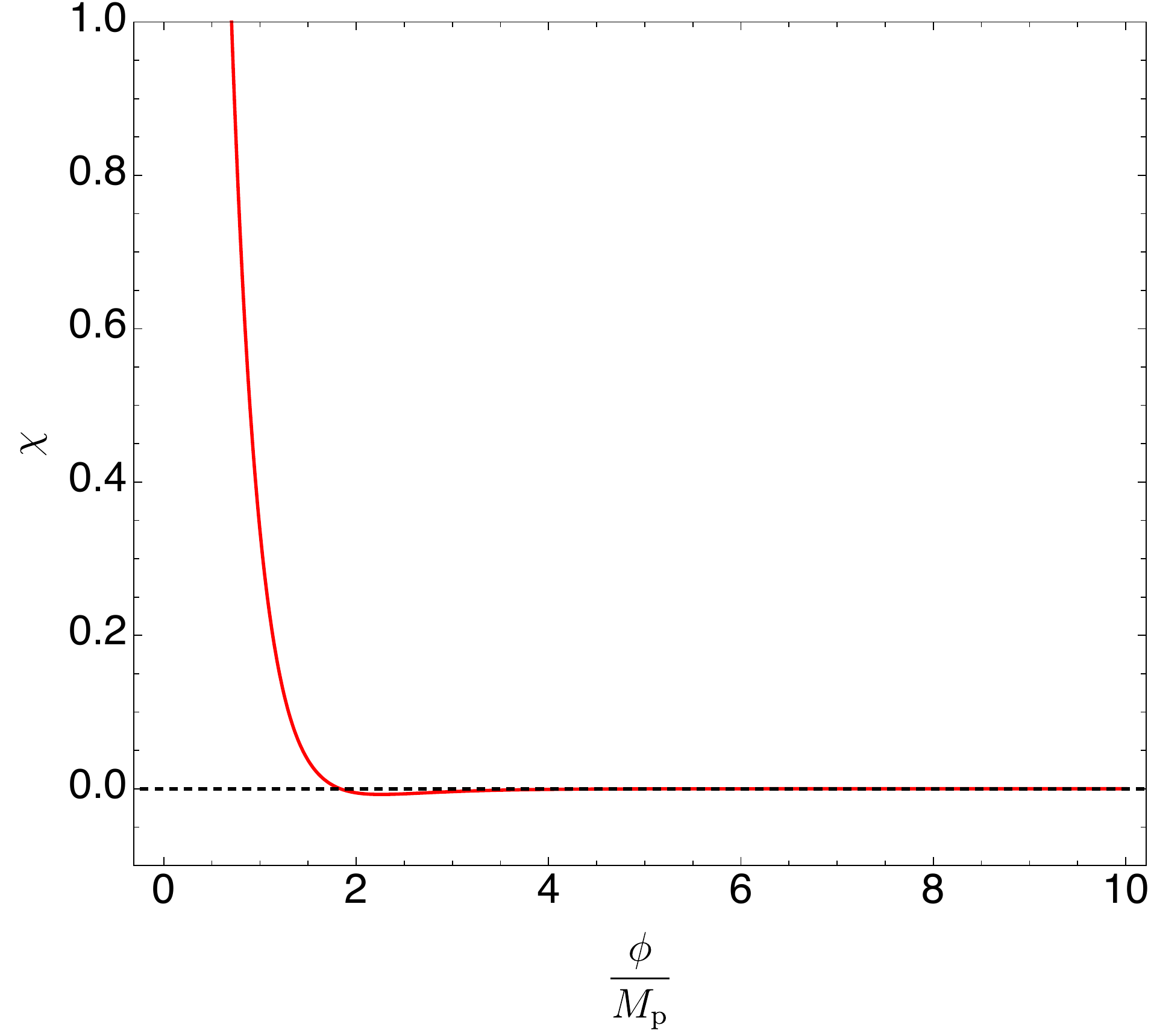}
	\caption{ a) The value of $\chi$ for $A=1$ and $B=0.2,0.5,1.0$. The SSWGC implies $\chi \geq 0$.
		b) The value of $\chi$ for the Starobinsky potential.}
	\label{starobinsky}
\end{figure}

We already mentioned that among polynomial potentials, the linear case is the only one that allows for trans-Planckian excursions.
So among chaotic inflation models \cite{chaotic}  the linear one is singled out as the unique class which can lead to sufficient inflation. 
As is well known, linear potentials may yield 50-60 e-folds of inflation with tensor perturbations with  $r\simeq 0.07$.
This relatively large value will soon be experimentally tested. Note that instead of a purely linear potential one may consider e.g.  the potential in
examples ii) or iii) above  which behave linearly for $|\phi|>M_{\text{p}}$. It is interesting to note that linear potentials do appear in string theory
realizations of monodromy inflation, see \cite{Silverstein:2008sg,McAllister:2008hb,McAllister:2014mpa,Dong:2010in,Marchesano:2014mla}.
There the stability of the potentials against corrections is guaranteed due to shift symmetries.
One type of potentials in this class has the form \cite{McAllister:2008hb}
\beq
V(b)\ =\  A(1\ +\ B\ b^2)^{1/2}
\eeq
where $b$ is a Type IIB (monodromic) axionic field and we set $M_{\text{p}}=1$. 
A simple way to check the validity of the Strong SWGC, Eq (\ref{ours}) is by plotting
\beq 
\chi\equiv 2\left(V^{\prime\prime\prime}\right)^{2}-\ V^{\prime\prime\prime\prime}V^{\prime\prime}-\  \left(\frac{V^{\prime\prime}}{M_\text{p}}\right)^2
\eeq
Then eq.(\ref{ours}) means $\chi\geq 0$.
We plot in fig.(\ref{starobinsky})-a that quantity for the above potential with $A=1$ and several values of $B$.
We see that the bound seems to be obeyed. In fact one can check that above $b\simeq 2$ the bound is slightly
violated at the per-mil label, something not visible in the figure. However we do not have control of the theory
to that level and one may say that this model passes the test.
There are several other schemes leading to
linear potentials which we will not discuss here, see  e.g.  \cite{McAllister:2008hb,Higgsotic}.

More generally one may consider monomial potentials of the form $V=\phi^a$, $a\geq 0$. For them the condition $\chi \geq 0$ gives $(a-1)(a-2) M^2_\text{p} -\phi^2 \geq 0$. For 
$0\leq a <1$ the potential has only tiny violations of the bound at small $\phi$, in the region $\phi < \sqrt{(a-1)(a-2)} M_\text{p} $.  The same formula applies for $a>2$, here the violations are large but are trans-Planckian for $a>2.7$.  However, 
for $1<a\leq2$ the bound is irremediably violated at all points of field space. Finally, $a=0$ and $a=1$ are the only pure monomials which satisfy the bound at all points of field space.

Another popular inflaton potential  is the Starobinsky model \cite{Starobinsky,Mukhanov}, which has the general form
\beq
V\ =\  \left( 1\ -\ e^{-\sqrt{2/3}\phi /M_{\text{p}}}\right)^ 2 \ .
\eeq
The same structure appears also in Higgs inflation \cite{higgsflation}. In Fig.(\ref{starobinsky})-b we plot $\chi$ as a function of the canonical field in units of Planck mass.
Essentially the same thing happens for the Starobinsky model. the simplest version of it would be  inconsistent with the Strong SWGC, since at some points in field space
 the condition is violated. It needs to be modified at large trans-Planckian distances. Adding a perturbation may possibly make it consistent.

 These are just a couple  of examples, just to show that the constraint is potentially very strong.
 It would be interesting to study these and other examples in more detail.

\subsection{Constraints on the SM from its 3D compactification}

Consider the SM compactified in a circle of radius $R$ down to 3D.  This radius is a modulus and has associated a quantum fluctuation field
$\phi$ with canonical kinetic term given by $R=r e^\frac{\phi}{M^{3d}_{\text{p}}\sqrt{2}}$. Here $r$ is any given reference scale to measure $R$ which we set equal to 1 GeV and $M^{3d}_{\text{p}}$ is the 3D Planck mass. Let us concentrate on the deep infrared region, well below the electron threshold,  with
$R\gg 1/m_e$. As explained in \cite{Nima}  the 3D one-loop effective potential for $R$ is given by the expression
   \begin{equation}
   V(R)\ =\ \ \frac {2\pi r^3 \Lambda_4}{R^2}\       -\ 4 \left( \frac {r^3}{720\pi R^6} \right) \ +\sum_{\nu_e,\nu_\mu,\nu_\tau}\ 
  r^{3} V_{\mathcal{C}}\left[R,m_{\nu_i}\right]\  .
   \label{potuno}
\end{equation}
The first term comes from the 4D cosmological term $\Lambda_4$ after
dimensional reduction (and going to the  3D Einstein frame). The second comes from the one-loop Casimir energy associated to the 
only two massless particles, the photon and the graviton. The factor 4 gives the number of helicity degrees of freedom of those fields.
The remaining term is the contribution to the Casimir energy of the three neutrinos compactified with periodic boundary conditions, and is given by
\begin{equation}
	V_{\mathcal{C}}\left[R,m_{\nu_i}\right]=\frac{n_{\nu_i}\ m_{\nu_i}^{2}}{8\pi^{4}R^{4}}\sum_{n=1}^{\infty}\frac{K_{2}(2\pi m_{\nu_i} n R )}{n^{2}} .
	\label{potdos}
\end{equation}
Here $n_{\nu_i}$ is the number of helicities for each neutrino (2 for Majorana and 4 for Dirac) and $K_{n}$ are modified Bessel functions of the second kind. This potential is reliable since
the contributions from higher thresholds are exponentially suppressed compared to the neutrino contributions by factors of order $e^{-m/m_\nu}$.
It has local minima in  AdS if neutrinos are Majorana \cite{Nima}. This is due to the fact that the lightest neutrino contributes positively to
the potential with 2 degrees of freedom, which is not enough to compensate for the 4 bosonic degrees of freedom contributing negatively from 
photon and graviton. However, if the lightest neutrino is Dirac (and it is lighter than the c.c. scale $\Lambda_4^{1/4}$) it contributes positively with 4
(instead of 2) degrees of freedom,   which is enough to compensate for the 4 massless degrees of freedom of the photon and the graviton. The potential is then
monotonously decreasing for large R and no AdS minima develops. This fact has been used to obtain bounds on the lightest neutrino mass and the 
4D cosmological constant \cite{IMV1} by imposing the condition suggested in \cite{OV} that AdS non-SUSY vacua are in the swampland
(see also \cite{2toro,Gonzalo,Hamada}).
  One obtains four very
relevant implications for the SM \cite{IMV1}:
\begin{itemize}

\item  The lightest neutrino is Dirac.

\item The lightest neutrino has a mass $m_{\nu_1}\leq 7.7\times 10^{-3}$ eV for normal hierarchy and
$m_{\nu_3}\leq 2.5 \times 10^{-3}$ eV for inverted hierarchy.

\item The 4D c.c. is bounded below by $\Lambda_4 \geq a(m_{\nu_1})^4$, with $a\simeq 1$. This is in agreement with the fact that the c.c. scale 
$\Lambda_4^{1/4}\simeq 10^{-3}$ eV is of order of the scale of neutrino masses.

\item Since Dirac neutrino masses are proportional to the Higgs vev (i.e. $m_{\nu_1}=h_{\nu_1}<H>$), an upper bound on the 
lightest neutrino mass implies un upper bound on the Higgs vev (at fixed Yukawa). This may give an understanding of the 
stability of the EW scale.

\end{itemize}

Here we will show that similar (but not identical)  interesting constraints on the SM may be obtained from the Strong SWGC here discussed if extended to 3D.
This is very attractive since, for the AdS swampland condition to apply, the AdS minima obtained must be absolutely stable, and this is
always difficult to prove (one cannot rule out some instability in the UV). It is important to remark that they are totally independent conjectures. In fact, the AdS criteria forbids Majorana masses while the Strong SWGC allows them. Interestingly, both set very similar bounds for the lightest Dirac neutrino mass. We will show is that unless the lightest Dirac neutrino is 
sufficiently light, the form of the scalar potential for $\sigma$ is not consistent with the 3D version of equation (\ref{ours}), for some value of
$R$ the scalar interaction becomes weaker than gravitation.

\begin{figure}
	\centering{}\includegraphics[scale=0.5]{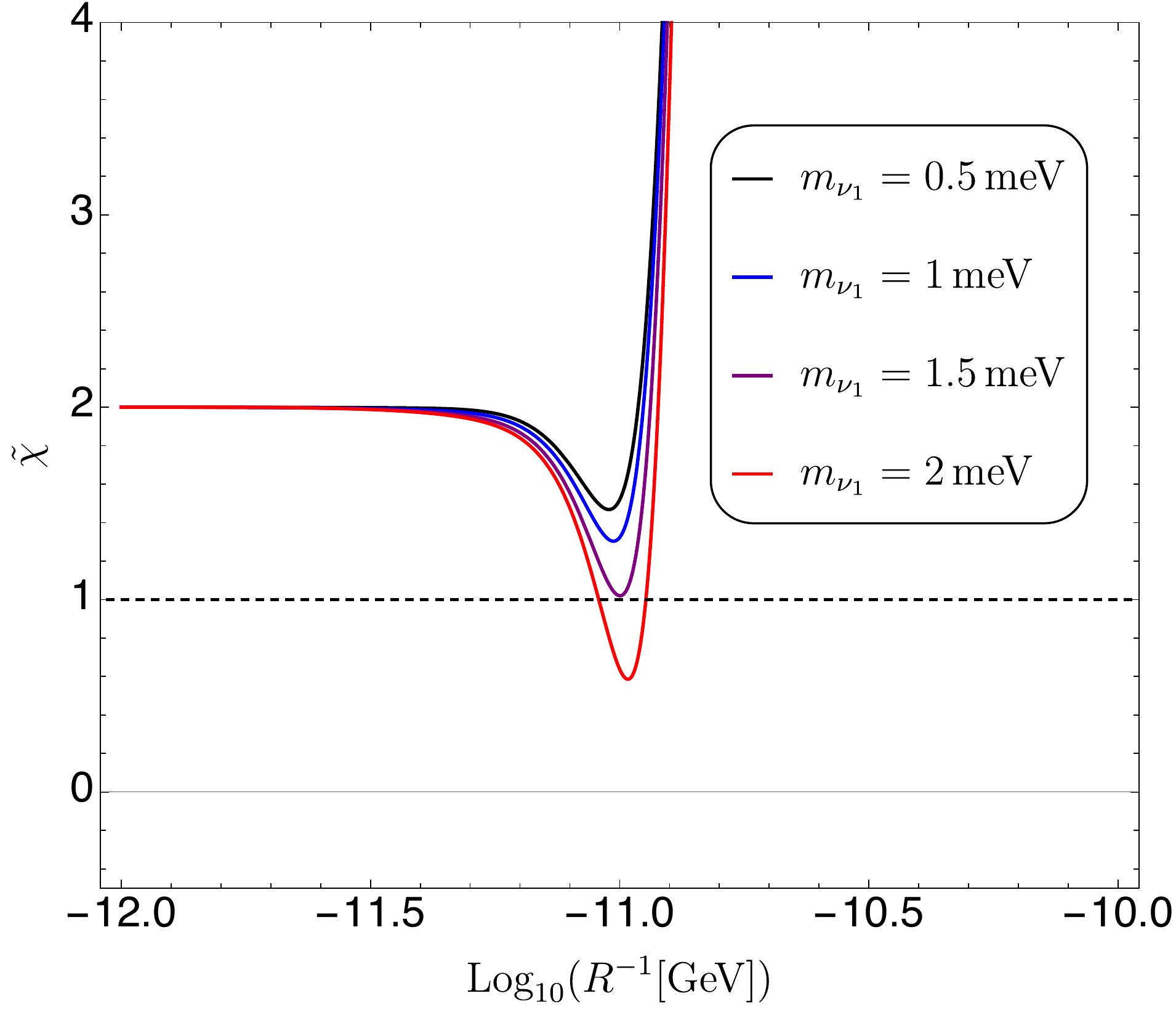}
	\caption{ Bound on neutrino mass for normal hierarchy}
	\label{neutrinos}
\end{figure}

We want to check if the effective potential of the SM compactified on a circle verifies Eq. (\ref{ours}). For practical reasons it is useful to define:
\begin{equation}
\frac{\tilde{\chi}}{M^2_\text{p}} \equiv2\left(\frac{V^{\prime\prime\prime}}{V^{\prime\prime}}\right)^{2}-\frac{V^{\prime\prime\prime\prime}}{V^{\prime\prime}},
\end{equation}
since the plots become easier to read. On the other hand, the intuition on what could change if a perturbation to the potential is included is lost, since we are taking ratios.  In terms of this new variable Eq. (\ref{ours}) is $\tilde{\chi} \geq 1$. In computing $\tilde{\chi}$ all derivatives are taken with respect to the canonical field $\phi$.
However, in Fig (\ref{neutrinos}) we plot it with respect to $R$, for simplicity.  The derivatives can be computed analytically using standard formulas involving the $K_{n}$ functions.
We find that for normal neutrino hierarchy the Strong SWGC is violated unless the lightest neutrino is lighter than $1.5\times 10^{-3}$ eV, see fig.(\ref{neutrinos}).
Interestingly,
in the case of inverted hierarchy we obtain a lower and not an upper bound on the lightest neutrino mass. In particular we find that the lightest neutrino must have $m_{\nu} \geq 1.6 \, \text{meV}$.
We can combine this bound with the results in \cite{IMV1,2toro} to conclude that, if both conjectures are true, the SM with inverse hierarchy would be in the Swampland.
Normal hierarchy is therefore another non-trivial prediction that arises from the conjecture. 
 It is interesting that present data already show a slight preference for the normal hierarchy.

\subsection{Constraints on the SM Higgs mass}

Independently of the above 3D constraints on the SM, one can consider possible SSWGC constraints directly in 4D.
Here the natural candidate to give rise to interactions weaker than gravity at some scale is the Higgs field.
The bound in eq(\ref{ours}) is a bound on the mass of a scalar (for all $\phi$). Since the mass is suppressed by a 
$1/M_{\text{p}}$ factor one may expect that it will trivially be obeyed by any particle physics model. In fact this is not necessarily the case.
It may happen that for some particular value of $\phi$ the term in the left hand side cancels exactly. In other words, defining
\beq
\delta(\phi)\ =\  
 2(V''')^2  -\ V''V''''\ \ ,
 \eeq
 one can obtain a bound  
 \beq
 m^4(\phi) \ \leq \  \delta (\phi) M_{\text{p}}^2 \ .
 \label{jerar}
 \eeq
 This means that if, {\it at some finite value of $\phi$}, $\delta(\phi)$ 
 vanishes or is very small, then the bound could be violated, indicating that our model is wrong or incomplete. 
 
 In the case of the
 physical Higgs field $H$ of the SM the above differential equations would have an additional  positive term  $(g_1^2+g_2^2)$ contributing to $\delta(H)$ from the exchange of
 electroweak gauge bosons. It is known 
  that above the EW region, the potential for the Higgs reaches a maximum at $Q_{\text{max}}$  and eventually 
 decreases and becomes negative in a region around $Q_{ins}\simeq 10^{11}-10^{13}$ GeV, see e.g.\cite{Espinosa} and references therein.  
 The maximum turns out to be close to the instability scale $Q_{ins}$ and $\delta(H)$ may  vanish close to that scale \cite{preparation}.
 This would 
  be the signal that either some new physics appear at that point or else one has to modify the SM below $Q_{\text{max}}$ so that
 this zero  of the Higgs interaction never appears. In particular, a SUSY version of the SM like e.g. the MSSM may avoid this 
 potential problem.  
 The Higgs potential in the SUSY case is monotonous, with no maxima at any intermediate region developing.
 Thus SUSY would be here present not to solve the hierarchy problem in the traditional sense (absence of 
 quadratic divergences) but rather to avoid that at any point the Higgs interaction becomes weaker than gravity.
 
  In this connection
 note that  in eq.(\ref{jerar}) the left hand side is quadratically divergent whereas the right hand side involves only logarithmically 
 divergent quantities. This would be indicating that the usual arguments about to the instability of scalar masses against quantum corrections 
 are at odds with constraints coming from WGC arguments.  An analogous observation but in a different context was already made in \cite{remmen}.
 We leave a detail  study of the numerical effect of our bound on the SM for future work
 \cite{preparation}.

\subsection{ Moduli fixing in flux  string vacua}

\begin{figure}
	\includegraphics[scale=0.4]{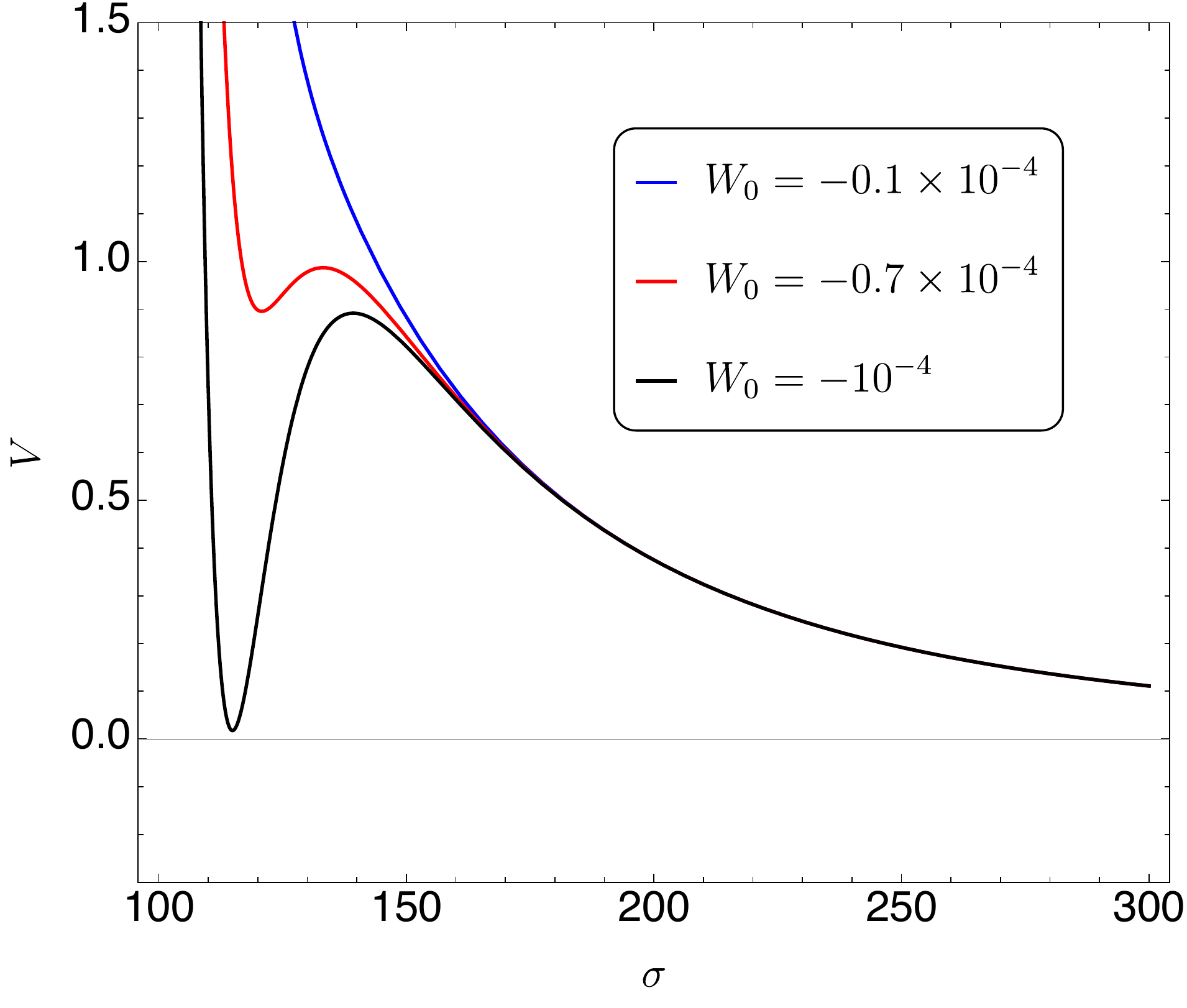}
	\includegraphics[scale=0.4]{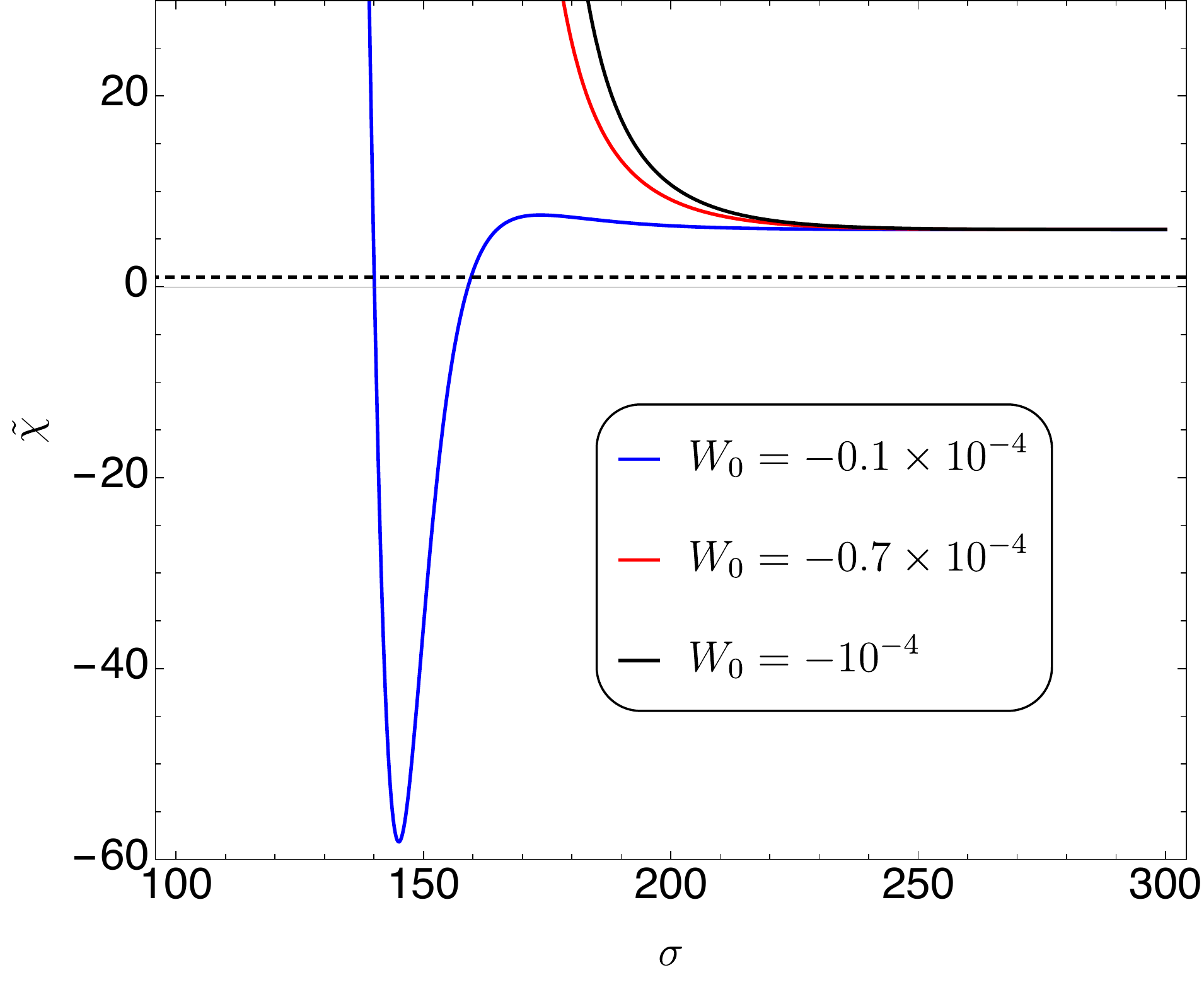}
	\caption{ KKLT potential for the parameters in the original ref.\cite{kklt} in black.  In red and blue we show alternative choices for $W_0$.
	For the choice in blue the bound is violated, but the potential does not have a minimum.}
	\label{KKLT}
\end{figure}

The  scalar potential of string compactification moduli is another instance in which interactions weaker than gravity could appear,
since moduli fields have Planck suppressed interactions. Let us consider here as the simplest example the KKLT \cite{kklt}.
In this model one assumes that the complex structure moduli are fixed due to fluxes at a higher scale.  One also assumes there
is a single Kahler modulus $T$ which also governs the strength of a gaugino condensation superpotential 
\beq
W\ =\ W_0\ +\ ce^{2\pi aT} \ .
\eeq
Here $W_0$ is a constant term induced by the fluxes and the gauge group resides on a set  of D7-branes. This yields a minimum in AdS.
In order to up-lift the vacuum to dS one assumes there are e.g. a set of anti-D3 branes on top of a throat at some point in the
compact CY. This yields an additional term $\delta V=D/(T+T^*)^3$, where D is proportional to the number of branes and may contain
model dependent suppression factors.
Setting the axion in $\text{Im} \, T$ to zero and letting $\text{Re}\,  T=\sigma$, the potential has the 
form
\beq
V_{KKLT}(\sigma)\ =\ 
\frac {\pi a c e^{-2\pi a\sigma} }{\sigma ^2}
\left( \frac {2\pi ac\sigma e^{-2\pi a \sigma} } {3}\ +\ W_0\ +\  ce^{-2\pi a \sigma }\right)\ +\ \frac {D}{8\sigma^3} \ .
\eeq
The kinetic term is 
\[
K_{i\overline{j}}\partial_{\mu}T^{i}\partial^{\mu}\overline{T}^{j}=\frac{3}{4T_{R}^{2}}\left(\left(\partial_{\mu}T_{R}\right)\left(\partial^{\mu}T_{R}\right)+\left(\partial_{\mu}T_{I}\right)\left(\partial^{\mu}T_{I}\right)\right) \ ,
\]
so the field is related to the canonically normalized field $\phi$ by $\sigma=T_{R}=e^{\sqrt{\frac{2}{3}} \phi}$.
The condition $\tilde{\chi}\geq 1 $ is given by a very complicated expression which is a
ratio of exponentials and polynomia in $T_{R}=\sigma$. We find that, as long as $W_{0}$ is large enough to generate a minimun, the potential verifies the SSWGC at all values of $\sigma$.
The form of the potential for the parameters given in  \cite{kklt} is shown  with a black line in fig.\ref{KKLT}.  The figure in the right
 shows the ratio 
$\tilde{\chi}=\delta(\sigma)/V''^2$ which should be everywhere bigger than one for the bound from eq.(\ref{ours}) to be verified.
One sees that the
bound is  respected for the original parameters in \cite{kklt}. However for smaller $ |W_0| $ (in blue in the figure) the bound may be violated,
although in the cases we have analyzed the potential has no minima.
At large $\sigma$ the largest exponential dominates and $\tilde{\chi}=6$ so the SSWGC is always verified.  It would be interesting to study 
the constraints in other moduli fixing string models.

\section{Discussion}

In this paper we have put forward a scalar version of the WGC. We call it {\it Strong SWGC} because we conjecture that it applies to 
any scalar, and not only to those which may be playing a role as ``WGC  scalars''. The conjectured is summarized by eq.(\ref{ours})
which should apply for all values of the field.
The constraint may be interpreted as the condition that the 
 strength of the interactions of any two scalars must be bigger than its gravitational interaction.
 This leads to a number of conclusions which unify and encompass some known swampland conjectures. The axion decay 
 constants are constrained by $f\leq M_{\text{p}}$. There are extremal solutions  leading  to an emergent dimension 
 with radius $R$ and masses with a structure $m^2\ =\ \frac {m_0^2}{1/(NR)^2\ +\ (R/M)^2} $,
  with a duality symmetry built in. The swampland distance conjecture arises 
 at small and large $R$ and requires the existence of extended objects (strings).
  There are two extra interesting results: i) There cannot be massive 
  scalars without any interaction other than gravity and ii) Among polynomial potentials only the linear one is consistent with the conditions
  and hence allow for trans-Planckian trips. 
 
The implications of this  SSWGC  are remarkable for both cosmology and particle physics: 1) In single field  chaotic inflation the linear
potential is uniquely selected as the only class of potentials in which trans-Planckian trips may take place. This leads to a solid prediction:
if large single field inflation is operative, the tensor-to-scalar ratio should be around $r=0.07$. Starobinsky or some linear monodromy 
inflation models need to be corrected to be viable 2) If applied to the 3D radion of the circle compactification of the SM, the SSWGC implies that the lightest
Dirac neutrino has a mass bounded above by the c.c. scale, $m_{\nu_1}\lesssim \Lambda_4^{1/4}$.  
Combining the results of \cite{IMV1,2toro} with the results of this work we find that the lightest neutrino must be Dirac and the
hierarchy must be normal (not inverted).  Furthermore,
the bound on the neutrino mass implies a constraint on the Higgs vev (for fixed Yukawa).  This would give an understanding of the Higgs
stability against quantum corrections in the SM.  Somewhat similar SM predictions were in fact already derived in terms of the AdS swampland conjecture of
\cite{OV}  in ref.\cite{IMV1}. However those predictions relied on the stability of the induced AdS 3D potential, which is difficult to establish in the
absence of UV information. In the derivation from the SSWGC here considered the information required is purely local and independent from any
UV information. 
3) The SSWGC may be applied to the Higgs field in the SM, suggesting that new physics should appear at an intermediate scale or
below. This would be independent of the traditional argument based on the absence of quadratic divergences, and 4) The SSWGC can be applied to
moduli fixing models of string compactification. The simplest KKLT scenario is consistent with the constraints, although the parameters 
of moduli fixing potentials would be constrained.

Although the conjecture looks very attractive and predictive and it is able to encompass several of the proposed swampland conjectures, 
further effort should be made to understand its physical origin as coming from a ``gravity as the weakest force" condition.
  In addition, the role of the extremal solutions as towers of solitonic states needs to be understood. 
  The generalization to more complex situations with many scalars is also important.
Finally, it would be interesting to find out what is it precisely that goes wrong when the scalar interaction is weaker than gravity.  
 While the WGC for charged particles and gauge bosons is relatively well understood in terms
 of extremal charged black-holes,  its generalization to scalar fields and interactions remains challenging. We hope the present paper may
 be useful to shed some new light into this question.

\newpage

\centerline{\bf \large Acknowledgments}

\bigskip

\noindent We thank J.R. Espinosa, A. Herr\'aez,  F. Marchesano, M. Montero, E. Palti,  A. Uranga, and I. Valenzuela for useful discussions. 
We thank Ido Ben-Dayan for prompting us to check the case of the SM Higgs potential.
This work has been supported by the ERC Advanced Grant SPLE under contract ERC-2012-ADG-20120216-320421, by the grants FPA2016-78645-P and
FPA2015-65480-P from the MINECO,  and the grant SEV-2016-0597 of the ``Centro de Excelencia Severo Ochoa" Programme.
 E.G. is supported by the Spanish FPU Grant No. FPU16/03985.


\vspace{1.3cm}


\begin{thebibliography}{99}

\bibitem{swampland}
	C.~Vafa, ``The String landscape and the swampland,''
	hep-th/0509212


\bibitem{WGC}
	N.~Arkani-Hamed, L.~Motl, A.~Nicolis and C.~Vafa,
	``The String landscape, black holes and gravity as the weakest force,''
	JHEP {\bf 0706} (2007) 060
	[hep-th/0601001]
	
	\bibitem{distance}
	H.~Ooguri and C.~Vafa,
	``On the Geometry of the String Landscape and the Swampland,''
	Nucl.\ Phys.\ B {\bf 766}, 21 (2007)
	[hep-th/0605264].

        \bibitem{review}
         E.~Palti,
  ``The Swampland: Introduction and Review,''
  arXiv:1903.06239 [hep-th].
        

	\bibitem{WGC1}
	T.~Rudelius,
	``Constraints on Axion Inflation from the Weak Gravity Conjecture,''
	JCAP {\bf 1509} (2015) no.09,  020
	[arXiv:1503.00795 [hep-th]]\\
	M.~Montero, A.~M.~Uranga and I.~Valenzuela,
	``Transplanckian axions!?,''
	JHEP {\bf 1508} (2015) 032
	[arXiv:1503.03886 [hep-th]]\\
	J.~Brown, W.~Cottrell, G.~Shiu and P.~Soler,
	``Fencing in the Swampland: Quantum Gravity Constraints on Large Field Inflation,''
	JHEP {\bf 1510} (2015) 023
	[arXiv:1503.04783 [hep-th]]\\
	J.~Brown, W.~Cottrell, G.~Shiu and P.~Soler,
	``On Axionic Field Ranges, Loopholes and the Weak Gravity Conjecture,''
	JHEP {\bf 1604}, 017 (2016)
	[arXiv:1504.00659 [hep-th]]\\
	B.~Heidenreich, M.~Reece and T.~Rudelius,
	``Weak Gravity Strongly Constrains Large-Field Axion Inflation,''
	JHEP {\bf 1512} (2015) 108
	[arXiv:1506.03447 [hep-th]]\\
	A.~de la Fuente, P.~Saraswat and R.~Sundrum,
	``Natural Inflation and Quantum Gravity,''
	Phys.\ Rev.\ Lett.\  {\bf 114} (2015) no.15,  151303
	[arXiv:1412.3457 [hep-th]]\\
	A.~Hebecker, P.~Mangat, F.~Rompineve and L.~T.~Witkowski,
	``Winding out of the Swamp: Evading the Weak Gravity Conjecture with F-term Winding Inflation?,''
	Phys.\ Lett.\ B {\bf 748} (2015) 455
	[arXiv:1503.07912 [hep-th]]\\
	T.~C.~Bachlechner, C.~Long and L.~McAllister,
	``Planckian Axions and the Weak Gravity Conjecture,''
	JHEP {\bf 1601} (2016) 091
	[arXiv:1503.07853 [hep-th]]\\
	T.~Rudelius,
	``On the Possibility of Large Axion Moduli Spaces,''
	JCAP {\bf 1504} (2015) no.04,  049
	[arXiv:1409.5793 [hep-th]]\\
	D.~Junghans,
	``Large-Field Inflation with Multiple Axions and the Weak Gravity Conjecture,''
	JHEP {\bf 1602} (2016) 128
	[arXiv:1504.03566 [hep-th]]\\
	K.~Kooner, S.~Parameswaran and I.~Zavala,
	``Warping the Weak Gravity Conjecture,''
	Phys.\ Lett.\ B {\bf 759}, 402 (2016)
	[arXiv:1509.07049 [hep-th]]\\
	D.~Harlow,
	``Wormholes, Emergent Gauge Fields, and the Weak Gravity Conjecture,''
	JHEP {\bf 1601}, 122 (2016)
	[arXiv:1510.07911 [hep-th]]\\
	L.~E.~Ib\'a\~nez, M.~Montero, A.~Uranga and I.~Valenzuela,
	``Relaxion Monodromy and the Weak Gravity Conjecture,''
	JHEP {\bf 1604} (2016) 020
	[arXiv:1512.00025 [hep-th]]\\
	A.~Hebecker, F.~Rompineve and A.~Westphal,
	``Axion Monodromy and the Weak Gravity Conjecture,''
	JHEP {\bf 1604} (2016) 157
	[arXiv:1512.03768 [hep-th]].
	
	
	
	
	\bibitem{WGC2} 
	B.~Heidenreich, M.~Reece and T.~Rudelius,
	``Evidence for a Lattice Weak Gravity Conjecture,''
	arXiv:1606.08437 [hep-th]\\
	M.~Montero, G.~Shiu and P.~Soler,
	``The Weak Gravity Conjecture in three dimensions,''
	arXiv:1606.08438 [hep-th]\\
	P.~Saraswat,
	``The Weak Gravity Conjecture and Effective Field Theory,''
	arXiv:1608.06951 [hep-th]\\
	D.~Klaewer and E.~Palti,
	``Super-Planckian Spatial Field Variations and Quantum Gravity,''
	arXiv:1610.00010 [hep-th]\\
	L.~McAllister, P.~Schwaller, G.~Servant, J.~Stout and A.~Westphal,
	``Runaway Relaxion Monodromy,''
	arXiv:1610.05320 [hep-th]\\
	A.~Herr\'aez and  L.~E.~Ib\'a\~nez,
	``An Axion-induced SM/MSSM Higgs Landscape and the Weak Gravity Conjecture,''
	JHEP {\bf 1702} (2017) 109
	[arXiv:1610.08836 [hep-th]] \\
	M.~Montero,
	``Are tiny gauge couplings out of the Swampland?,''
	[arXiv:1708.02249 [hep-th]]\\
	L.~E.~Ib\'a\~nez and M.~Montero,
	``A Note on the WGC, Effective Field Theory and Clockwork within String Theory,''
	 JHEP {\bf 1802} (2018) 057
  [arXiv:1709.02392 [hep-th]]\\
	G.~Aldazabal and L.~E.~Ib\'a\~nez
    `A Note on 4D Heterotic String Vacua, FI-terms and the Swampland,''
  Phys.\ Lett.\ B {\bf 782} (2018) 375
   [arXiv:1804.07322 [hep-th]].
	
		
\bibitem{WGC3}
 C.~Cheung, J.~Liu and G.~N.~Remmen,
  ``Proof of the Weak Gravity Conjecture from Black Hole Entropy,''
arXiv:1801.08546 [hep-th]\\
 T.~W.~Grimm, E.~Palti and I.~Valenzuela,
  \``Infinite Distances in Field Space and Massless Towers of States,''
arXiv:1802.08264 [hep-th]\\
  B.~Heidenreich, M.~Reece and T.~Rudelius,
  ``Emergence and the Swampland Conjectures,''
arXiv:1802.08698 [hep-th]\\
 S.~Andriolo, D.~Junghans, T.~Noumi and G.~Shiu,
  ``A Tower Weak Gravity Conjecture from Infrared Consistency,''
arXiv:1802.04287 [hep-th]\\
 R.~Blumenhagen, D.~Klaewer, L.~Schlechter and F.~Wolf,
  ``The Refined Swampland Distance Conjecture in Calabi-Yau Moduli Spaces,''
arXiv:1803.04989 [hep-th]\\
 A.~Landete and G.~Shiu,
  ``Mass Hierarchies and Dynamical Field Range,''
  arXiv:1806.01874 [hep-th]\\
   Y.~Hamada, T.~Noumi and G.~Shiu,
  ``Weak Gravity Conjecture from Unitarity and Causality,''
  arXiv:1810.03637 [hep-th].
	
\bibitem{timo}
 S.~J.~Lee, W.~Lerche and T.~Weigand,
  ``Tensionless Strings and the Weak Gravity Conjecture,''
  JHEP {\bf 1810} (2018) 164
  [arXiv:1808.05958 [hep-th]]\\
    S.~J.~Lee, W.~Lerche and T.~Weigand,
  ``Modular Fluxes, Elliptic Genera, and Weak Gravity Conjectures in Four Dimensions,''
  arXiv:1901.08065 [hep-th].
  



	\bibitem{vafafederico}
	T.~D.~Brennan, F.~Carta and C.~Vafa,
	``The String Landscape, the Swampland, and the Missing Corner,''
	arXiv:1711.00864 [hep-th].	
	
	\bibitem{Palti}
  E.~Palti,
  ``The Weak Gravity Conjecture and Scalar Fields,''
  JHEP {\bf 1708} (2017) 034
  [arXiv:1705.04328 [hep-th]].

\bibitem{timoscalar}
 S.~J.~Lee, W.~Lerche and T.~Weigand,
  ``A Stringy Test of the Scalar Weak Gravity Conjecture,''
  Nucl.\ Phys.\ B {\bf 938} (2019) 321
  [arXiv:1810.05169 [hep-th]].



\bibitem{Lust}
  D.~Lust and E.~Palti,
  ``Scalar Fields, Hierarchical UV/IR Mixing and The Weak Gravity Conjecture,''
  JHEP {\bf 1802} (2018) 040
   [arXiv:1709.01790 [hep-th]].


\bibitem{paltidistance}
 D.~Klaewer and E.~Palti,
 ``Super-Planckian Spatial Field Variations and Quantum Gravity,''
  JHEP {\bf 1701} (2017) 088
  [arXiv:1610.00010 [hep-th]].


	
	\bibitem{dS1}
  G.~Obied, H.~Ooguri, L.~Spodyneiko and C.~Vafa,
  ``De Sitter Space and the Swampland,''
  arXiv:1806.08362 [hep-th].
	
	\bibitem{dS2}
  P.~Agrawal, G.~Obied, P.~J.~Steinhardt and C.~Vafa,
  ``On the Cosmological Implications of the String Swampland,''
  Phys.\ Lett.\ B {\bf 784} (2018) 271
  [arXiv:1806.09718 [hep-th]].
  
\bibitem{krishnan}
  S.~K.~Garg and C.~Krishnan,
  ``Bounds on Slow Roll and the de Sitter Swampland,''
  arXiv:1807.05193 [hep-th].
  
  \bibitem{dS3}
  H.~Ooguri, E.~Palti, G.~Shiu and C.~Vafa,
  ``Distance and de Sitter Conjectures on the Swampland,''
  arXiv:1810.05506 [hep-th].
	
	
		\bibitem{masdS}
	 G.~Dvali and C.~Gomez,
  ``On Exclusion of Positive Cosmological Constant,''
  arXiv:1806.10877 [hep-th].\\
	 G.~Dvali, C.~Gomez and S.~Zell,
  ``Quantum Breaking Bound on de Sitter and Swampland,''
  arXiv:1810.11002 [hep-th]\\
	 D.~Andriot,
  ``On the de Sitter swampland criterion,''
  Phys.\ Lett.\ B {\bf 785} (2018) 570
  [arXiv:1806.10999 [hep-th]]\\
   C.~Roupec and T.~Wrase,
  ``de Sitter extrema and the swampland,''
  Fortsch.\ Phys.\  {\bf 2018} 1800082
    [arXiv:1807.09538 [hep-th]]\\
   J.~P.~Conlon,
  ``The de Sitter swampland conjecture and supersymmetric AdS vacua,''
  Int.\ J.\ Mod.\ Phys.\ A {\bf 33} (2018) no.29,  1850178
  [arXiv:1808.05040 [hep-th]]\\
    S.~Kachru and S.~P.~Trivedi,
  ``A comment on effective field theories of flux vacua,''
  arXiv:1808.08971 [hep-th]\\
	 H.~Murayama, M.~Yamazaki and T.~T.~Yanagida,
  ``Do We Live in the Swampland?,''
  arXiv:1809.00478 [hep-th]\\
   G.~Buratti, E.~Garcia-Valdecasas and A.~M.~Uranga,
  ``Supersymmetry Breaking Warped Throats and the Weak Gravity Conjecture,''
  arXiv:1810.07673 [hep-th]\\
    M.~Montero,
  ``A Holographic Derivation of the Weak Gravity Conjecture,''
  arXiv:1812.03978 [hep-th]\\
   C.~Cordova, G.~B.~De Luca and A.~Tomasiello,
  ``Classical de Sitter Solutions of Ten-Dimensional Supergravity,''
  arXiv:1812.04147 [hep-th]\\
   G.~Buratti, J.~Calderon and A.~M.~Uranga,
  ``Transplanckian Axion Monodromy !?,''
  arXiv:1812.05016 [hep-th].
	

\bibitem{irene}
  T.~W.~Grimm, E.~Palti and I.~Valenzuela,
  ``Infinite Distances in Field Space and Massless Towers of States,''
  JHEP {\bf 1808} (2018) 143
   [arXiv:1802.08264 [hep-th]].

	\bibitem{thomas}
  T.~W.~Grimm, C.~Li and E.~Palti,
  ``Infinite Distance Networks in Field Space and Charge Orbits,''
  arXiv:1811.02571 [hep-th].


	
	 
\bibitem{chaotic}
  A.~D.~Linde,
  {\em ``Chaotic Inflation,''}
  Phys.\ Lett.\ B {\bf 129} (1983) 177.

    

	
	
\bibitem{Silverstein:2008sg} 
  E.~Silverstein and A.~Westphal,
  {\em ``Monodromy in the CMB: Gravity Waves and String Inflation,''}
  Phys.\ Rev.\ D {\bf 78}, 106003 (2008)
  [arXiv:0803.3085 [hep-th]].


\bibitem{McAllister:2008hb} 
  L.~McAllister, E.~Silverstein and A.~Westphal,
  {\em ``Gravity Waves and Linear Inflation from Axion Monodromy,''}
  Phys.\ Rev.\ D {\bf 82}, 046003 (2010)
  [arXiv:0808.0706 [hep-th]].

\bibitem{McAllister:2014mpa} 
  L.~McAllister, E.~Silverstein, A.~Westphal and T.~Wrase,
  ``The Powers of Monodromy,''
  JHEP {\bf 1409}, 123 (2014)
  [arXiv:1405.3652 [hep-th]].



\bibitem{Dong:2010in}
  X.~Dong, B.~Horn, E.~Silverstein and A.~Westphal,
  {\em ``Simple exercises to flatten your potential,''}
  Phys.\ Rev.\ D {\bf 84} (2011) 026011
  [arXiv:1011.4521 [hep-th]].

\bibitem{Marchesano:2014mla} 
  F.~Marchesano, G.~Shiu and A.~M.~Uranga,
  {\em ``F-term Axion Monodromy Inflation,''}
  JHEP {\bf 1408}, 157 (2014)
  [arXiv:1406.2729 [hep-th]].


\bibitem{baumann} 
  D.~Baumann and L.~McAllister,
  {\em ``Inflation and String Theory,''}
  arXiv:1404.2601 [hep-th].


   
\bibitem{Higgsotic}
  L.~E.~Ib\'a\~nez and I.~Valenzuela,
  {\em ``The Higgs Mass as a Signature of Heavy SUSY,''}
  JHEP {\bf 1305} (2013) 064
  [arXiv:1301.5167 [hep-ph]]\\
  \  L.~E.~Ib\'a\~nez, F.~Marchesano and I.~Valenzuela,
  ``Higgs-otic Inflation and String Theory,''
  JHEP {\bf 1501} (2015) 128
  doi:10.1007/JHEP01(2015)128
  [arXiv:1411.5380 [hep-th]].
  
  \bibitem{Starobinsky}
  A.~A.~Starobinsky,
  ``A New Type of Isotropic Cosmological Models Without Singularity,''
  Phys.\ Lett.\ B {\bf 91} (1980) 99
   [Phys.\ Lett.\  {\bf 91B} (1980) 99]
   [Adv.\ Ser.\ Astrophys.\ Cosmol.\  {\bf 3} (1987) 130].
   
   
   \bibitem{Mukhanov}
  V.~F.~Mukhanov and G.~V.~Chibisov,
  ``Quantum Fluctuations and a Nonsingular Universe,''
  JETP Lett.\  {\bf 33} (1981) 532
   [Pisma Zh.\ Eksp.\ Teor.\ Fiz.\  {\bf 33} (1981) 549].
   
   
  	
  \bibitem{higgsflation}
  For a review and references see 
   F.~Bezrukov,
  {\em ``The Higgs field as an inflaton,''}
  Class.\ Quant.\ Grav.\  {\bf 30} (2013) 214001
  [arXiv:1307.0708 [hep-ph]].
 
	
	\bibitem{Nima}
	N.~Arkani-Hamed, S.~Dubovsky, A.~Nicolis and G.~Villadoro,
	``Quantum Horizons of the Standard Model Landscape,''
	JHEP {\bf 0706} (2007) 078
	[hep-th/0703067 [HEP-TH]].
	

	\bibitem{OV}
	H.~Ooguri and C.~Vafa,
	``Non-supersymmetric AdS and the Swampland,''
    Adv.\ Theor.\ Math.\ Phys.\  {\bf 21} (2017) 1787
  [arXiv:1610.01533 [hep-th]].
			
		\bibitem{IMV1}
	L.~E.~Ib\'a\~nez, V.~Martin-Lozano and I.~Valenzuela,
	``Constraining Neutrino Masses, the Cosmological Constant and BSM Physics from the Weak Gravity Conjecture,''
	JHEP {\bf 1711} (2017) 066  [arXiv:1706.05392 [hep-th]].
	
	
	\bibitem{OV}
	H.~Ooguri and C.~Vafa,
	``Non-supersymmetric AdS and the Swampland,''
    Adv.\ Theor.\ Math.\ Phys.\  {\bf 21} (2017) 1787
  [arXiv:1610.01533 [hep-th]].

	
		
\bibitem{2toro}
E.~Gonzalo, A.~Herr\'aez and L.~E.~Ib\'a\~nez,
``AdS-phobia, the WGC, the Standard Model and Supersymmetry,''
JHEP {\bf 1806} (2018) 051
[arXiv:1803.08455 [hep-th]].
	

\bibitem{Gonzalo}
  E.~Gonzalo and L.~E. Ib\'a\~nez,
  ``The Fundamental Need for a SM Higgs and the Weak Gravity Conjecture,''
  Phys.\ Lett.\ B {\bf 786} (2018) 272
    [arXiv:1806.09647 [hep-th]].
	
	
	
	\bibitem{Hamada}
	Y.~Hamada and G.~Shiu,
	``Weak Gravity Conjecture, Multiple Point Principle and the Standard Model Landscape,''
	JHEP {\bf 1711} (2017) 043
	[arXiv:1707.06326 [hep-th]].

\bibitem{preparation}
J.R. Espinosa, E. Gonzalo and L.E. Ib\'a\~nez, in progress (2019).


	
	\bibitem{Espinosa}
  G.~Degrassi, S.~Di Vita, J.~Elias-Miro, J.~R.~Espinosa, G.~F.~Giudice, G.~Isidori and A.~Strumia,
  ``Higgs mass and vacuum stability in the Standard Model at NNLO,''
  JHEP {\bf 1208} (2012) 098
  doi:10.1007/JHEP08(2012)098
  [arXiv:1205.6497 [hep-ph]].
  
  
  \bibitem{remmen}
  C.~Cheung and G.~N.~Remmen,
	``Naturalness and the Weak Gravity Conjecture,''
	Phys.\ Rev.\ Lett.\  {\bf 113} (2014) 051601
	[arXiv:1402.2287 [hep-ph]]\\

	
\bibitem{kklt}
  S.~Kachru, R.~Kallosh, A.~D.~Linde and S.~P.~Trivedi,
  ``De Sitter vacua in string theory,''
  Phys.\ Rev.\ D {\bf 68} (2003) 046005
   [hep-th/0301240].
     
     
	
	\bibitem{Blumenhagen:2019qcg}
  R.~Blumenhagen, D.~Klaewer and L.~Schlechter,
  ``Swampland Variations on a Theme by KKLT,''
  arXiv:1902.07724 [hep-th].
	
	
				
	
\end{thebibliography}
\end{document}